\documentstyle[preprint,eqsecnum,aps,epsfig]{revtex}

\begin{document}

\title{Effects of neutrino temperatures and mass hierarchies
on the detection of supernova neutrinos}

\author{Shao-Hsuan Chiu\thanks{chiu@physics.purdue.edu}
 and T. K. Kuo\thanks{tkkuo@physics.purdue.edu}}

\address{Department of Physics, Purdue University, West Lafayette, IN 47907}

\maketitle

\newif\iftightenlines\tightenlinesfalse
\tightenlines\tightenlinestrue

\begin{abstract}

Possible outcomes of neutrino events at both Super-Kamiokande and
SNO for a type-II supernova are analyzed considering
the uncertainties in SN neutrino spectra (temperature)
at emission, which may
complicate the interpretation of the observed events.
With the input of parameters deduced from the current
solar and atmospheric experiments, consequences of
direct-mass hierarchy $m_{\nu_{\tau}} \gg m_{\nu_{\mu}} > m_{\nu_{e}}$
and inverted-mass hierarchy 
$m_{\nu_{e}} > m_{\nu_{\mu}} \gg m_{\nu_{\tau}}$ are
investigated.
Even if the $\nu$ temperatures are not precisely known, we found that 
future experiments are likely to be able to separate the
currently accepted solutions to the solar neutrino problem (SNP):
large angle MSW,
small angle MSW, and the vacuum oscillation,
as well as to distinguish between the direct and inverted mass
hierarchies of the neutrinos.

\end{abstract}




\section{Introduction}

During the past few decades, elaborate solar neutrino~\cite{davis:94}
and atmospheric neutrino~\cite{kk:94} experiments
 have provided a wealth of convincing
evidences for the existence of massive neutrinos and neutrino mixing,
which could have an essential impact on
particle physics, astrophysics
and cosmology.  Attentions have been focused on solving the puzzles of
unexpected discrepancies between calculated and observed neutrino
fluxes.  
Instead of the
more difficult and unlikely solution from an improved solar 
model~\cite{bahcall:92},
the solar $\nu_{e}$ deficit could be reconciled with
the prediction if
neutrino oscillations occur either in vacuum or in the
 presence of solar matter.  

The flavor oscillation can be parameterized by
the mass-squared differences of the neutrino mass eigenstates
$\Delta m^{2} \equiv m_{i}^{2}-m_{j}^{2} \,(i,j= 1,2,3)$ and
$\theta_{ij}$,
the mixing angles between weak eigenstates and mass
eigenstates of the neutrinos ($\theta_{ij}\leq \frac{\pi}{4}$ is assumed). 
 In terms of these parameters,
the just-so vacuum oscillation~\cite{glashow:99} requires
$6 \times 10^{-11} \leq \Delta m^{2} \leq 60 \times 10^{-11}$ eV$^{2}$ and
 $\sin^{2}2\theta \simeq 1$, while the
MSW resonant 
effect~\cite{msw:7885} in the Sun
becomes important if $4\times 10^{-6}$ eV$^{2}$
$\leq \Delta m^{2} \leq 7\times 10^{-5}$ eV$^{2}$, $\sin^{2}2\theta
\simeq 0.6-0.9 $ (large angle solution), or  $3\times 10^{-6}$ eV$^{2}$
$ \leq \Delta m^{2} \leq 12\times 10^{-6}$ eV$^{2}$, $0.003 \leq
\sin^{2}2\theta \leq 0.01$ (small angle solution)~\cite{teshima:99}.
Recent atmospheric neutrino data from the Super-Kamiokande~\cite{fukuda:98}
further provide a strong evidence in support of neutrino oscillation
as the cause to deficit of muon neutrinos, 
provided $\Delta m^{2} \sim 10^{-2}-10^{-3}$ eV$^{2}$
and $\sin^{2} 2\theta > 0.82$.  It is clear that
this solution to the neutrino anomaly in the atmosphere
represents quite a distinct
area in the parameter space as compared to that of the solar neutrino
deficit.  Based on
 the conclusive LEP experiment at CERN~\cite{data:96}
 that there are three flavors of
light, active neutrinos participating in the weak interaction, a
direct-mass hierarchy $m_{\nu_{\tau}} \gg m_{\nu_{\mu}} > m_{\nu_{e}}$
, with $\Delta^{2}_{32} \simeq
\Delta^{2}_{31} \gg \delta^{2}_{21}$
( $\Delta^{2}_{32} \equiv m^{2}_{3}-m^{2}_{2}$
and $\delta^{2}_{21} \equiv m^{2}_{2}-m^{2}_{1}$)
 naturally accommodates the scales of both the two mass-squared differences and
provides solutions to both
puzzles: the conversion $\nu_{e} \rightarrow \nu_{\mu}$ causes the
 observed deficit in the solar $\nu_{e}$ flux and the vacuum
oscillation $\nu_{\mu} \rightarrow
\nu_{\tau}$ suppresses the $\nu_{\mu}$ flux in the atmosphere.

In addition to the Sun and the atmosphere, type-II supernovae are
also natural sources that emit neutrinos.
Despite the first-ever
observation of SN neutrino signals from
SN 1987A~\cite{sn:87}, detailed neutrino spectral shapes
have not yet been determined
with certainty due to low statistics and the physical processes that are
not well understood.
 This difficulty is accompanied by, for instance, the uncertainties
 in the characteristic temperatures
$T_{\nu}$ as neutrinos were 
 emitted from the neutrino-spheres.   Consequently, the
interpretation of future measurements of SN neutrinos would contain ambiguity
in that the observed
spectrum, which may have been deformed through conversion processes, could be
simulated by different set of parameters at different temperatures.

It is therefore worthwhile to investigate how the uncertainty in $T_{\nu}$ 
could impact the interpretation of events at 
terrestrial detectors.  
In this paper, the parameters that solve solar and atmospheric neutrino
problems are taken as inputs, a natural choice 
as also adopted by some earlier 
works~\cite{dutta:99}~\cite{dighe:99}. 
In addition, with the uncertainty in $T_{\nu}$ considered, we 
study whether a particular set of parameters could be singled out 
by future observations of SN neutrinos.

Unlike solar neutrinos, the initial neutrino flux
from a supernova contains all flavors
of neutrino: $\nu_{e}, \nu_{\mu},\nu_{\tau}$ and
their anti-particles.
Under the direct-mass hierarchy of neutrinos,
the original $\nu$ spectra will 
be modified by the MSW
effect as neutrinos propagate through the resonance.
The $\overline{\nu}$ spectra, on the contrary, is subject only to vacuum
oscillation which yields an large averaged survival probability of
$\overline{\nu}_{e}: P(\overline{\nu}_{e} \rightarrow
\overline{\nu}_{e}) \geq \frac{1}{2}$.
The high-energy $\overline{\nu}_{\mu}
(\overline{\nu}_{\tau})$ would not be converted to the easily
detectable $\overline{\nu}_{e}$ (for instance, at Super-Kamiokande)
through the MSW effect
 unless neutrino masses are inverted,
in which case the heavier
mass eigenstate has a larger component in $\nu_{e}$ than in $\nu_{\mu}$
or $\nu_{\tau}$. 
Since the mixing angles are defined
in the first octant, the weak eigenstates $\nu_{\tau}, \nu_{\mu}$, and
$\nu_{e}$ are predominant in the mass eigenstates $\nu_{3}, \nu_{2}$, and
$\nu_{1}$, respectively.  Under the direct-mass hierarchy where  
$m_{\nu_{\tau}} > m_{\nu_{\mu}} > m_{\nu_{e}}$, the mass eigenstates
follow the hierarchy $m_{3} > m_{2} > m_{1}$, while in the inverted-mass
hierarchy, for instance, $m_{\nu_{e}} > m_{\nu_{\mu}} > m_{\nu_{\tau}}$,
the pattern $m_{1} > m_{2} > m_{3}$ follows.
 Some models and phenomenological consequences involving inverted
 neutrino masses have been discussed~\cite{silk:96}. 
Although current MSW solutions to the solar neutrino problem (SNP) 
have excluded $\nu_{e}$ as the heavier eigenstate, the inverse hierarchy
could remain viable if the just-so vacuum oscillation is the 
solution for the SNP.
If the inverted masses do apply and the resonance conditions
for the anti-neutrinos are met, this could lead to
an effective conversion
between $\overline{\nu}_{e}$ and the higher-energy
$\overline{\nu}_{\mu}(\overline{\nu}_{\tau})$ to yield
copious $\overline{\nu}_{e}$-type events in the earth-bound detector.
With the uncertainty in $T_{\nu}$ considered, it is our second goal 
to investigate influences of both the direct and inverted mass
hierarchies to future observations of SN neutrinos 
and how future measurements can play a role in this unsettled
issue of direct versus inverted neutrino masses.

This paper is organized as follows.
In section II  we summarize the general features of stellar collapse
and properties of the emitted neutrinos, and show how the
uncertainty in neutrino temperature 
could affect the outcomes in the detector.
Section III and Section IV contain more general results expected
from the future observations
at both Super-Kamiokande and SNO, for direct and inverted masses, respectively.
Based on the measurements,  possible schemes
which could provide discrimination among input parameters
and between the two mass hierarchies are proposed.
 Section V contains discussions and our concluding remarks.

\section{Detection of supernova neutrinos}
\subsection{SN neutrinos and neutrino parameters}

A massive star ($M \geq 8 M_{\odot}$)
becomes unstable at the last stage of its evolution.  When the mass of the
iron core reaches the Chandrasekhar limits ($\sim 1.4 M_{\odot}$),
it begins to collapse into a compact object of extremely high density,
and the gravitational binding energy is released in the form of neutrinos.
Mayle \emph{et al.}~\cite{mayle:87} have pointed out that the total
emitted energy, the averaged neutrino luminosity, and the mean neutrino
energy are independent of the explosive mechanism but depend only
 on the mass of the initial iron core.  Regardless of the details of
collapsed and bounce, it is well established that to form a typical
neutron star after the collapse, an amount of
$\sim 3\times 10^{53}$ erg, about 99\% of the binding energy would be
released in the form of neutrinos.  Each (anti)neutrino species will carry
away about the same amount of energy.

Neutrinos are emitted from a collapsed star through two different processes:
 neutronization burst during the pre-bounce phase and thermal emission in the
post-bounce phase.  The neutronization burst of a $\nu_{e}$
flux is produced by the
electron capture on protons:
$e^{-}+p\rightarrow n+\nu_{e}$.  The thermal emission creates
$\nu \overline{\nu}$ pairs of all three flavors via the annihilation of
$e^{+}e^{-}$ pairs: $e^{+}e^{-}\rightarrow \nu_{\ell}+\overline{\nu}_{\ell} \,
 (\ell=e,\mu,\tau)$.  The duration of neutronization burst lasts about a few
millisecond and takes away 1\%-10\% of the total binding energy.  The thermal
emission phase has a much wider spread of time structure,
in the order of 10 seconds.

The initial neutrino spectrum is usually approximated by a Fermi-Dirac
or a Boltzmann distribution with a constant temperature and zero chemical
potential.  To reduce the high-energy tail of the Fermi-Dirac distribution,
some elaborate models introduce a nonzero
chemical potential~\cite{wilson:98}.
It is clear that the event numbers in a detector depend crucially on the
$\nu$ temperature.  However,
the numerical calculations based upon various models and physical arguments
give rise to relatively wide ranges of temperature for
each neutrino species~\cite{schramm:87} and this uncertainty in $T_{\nu}$
could complicate the signatures concerning the oscillation of neutrinos from
a supernova.

One may refer to Ref.~\cite{kuo:89} for a review of neutrino oscillations. 
Although all the three flavors are emitted from a supernova, the phenomenon
of SN neutrino oscillation can be well described through $P(\nu_{e}
\rightarrow \nu_{e})$ and $P(\overline{\nu}_{e} \rightarrow
\overline{\nu}_{e})$~\cite{kuo:88}.
Under the direct-mass hierarchy, the probability
$P(\overline{\nu}_{e} \rightarrow \overline{\nu}_{e})$ is nearly
independent of
energy and is approximated by the vacuum oscillation expression.
The probability
$P(\nu_{e} \rightarrow \nu_{e})$ is energy-dependent and contains four
parameters under a proper parameterization of the mixing matrix
in the full 3-$\nu$ formalism:
$\delta_{21}^{2}, \Delta_{32}^{2} \simeq \Delta_{31}^{2},
\theta_{21}, \phi_{31}$.  In what
follows, these four parameters will simply be denoted as $\delta^{2}, \Delta^{2},
\theta$, and $\phi$, respectively.

Among the above four parameters, the angle $\phi$ is special in certain aspect.
In addition to the limit $\phi<12^{\circ}$ at 90\% C.L. set by the
CHOOZ~\cite{chooz:98} long baseline reactor in the
disappearance mode $\overline{\nu}_{e} \rightarrow \overline{\nu}_{x}$,
an analysis in Ref.~\cite{teshima:99}
also has given allowed ranges of $\delta^{2}$ and $ \sin^{2}2\theta$ for
 $\phi<20^{\circ}$.  Note that to zeroth order of
$\delta^{2}/\Delta^{2}$, the probability becomes~\cite{fogli:96}
  \begin{equation}
   P(\nu_{e} \rightarrow \nu_{e}) \simeq \cos^{4}\phi P_{2\nu}+\sin^{4}\phi. 
  \end{equation}
One may examine 
$\phi$ in more details through the iso-probability contours for
SN neutrinos at several
distinct scales of $\delta^{2}/E_{\nu}$, as shown in Figure 1.
Within the interested range of $\theta$,
( $\sin^{2}2\theta\geq 0.003$, or $\log_{10} \tan^{2}
2\theta \geq -2.5$ ), the $P(\nu_{e} \rightarrow \nu_{e})$ contours
 are almost independent of $\phi$ for $\phi < 12^{\circ}$
( $ \log_{10}\tan^{2}2\phi < -0.7$ ).  The parameter $\phi$
begins to show slight influence on
$P(\nu_{e} \rightarrow \nu_{e})$ contours only at very small $\theta$
and very large $\phi$.
Hence, for our purpose the choice of $\phi$ within $\phi<20^{\circ}$
would only affect the results slightly.  The value
$\phi=10^{\circ}$ will be adopted for definiteness. 
As for other
input parameters, the following are taken: 

Large angle MSW solution(LA): 
$\delta^{2} = 10^{-5}$ eV$^{2}, \sin^{2} 2\theta = 0.75.$ 

Small angle MSW solution(SA):  
$\delta^{2} = 6\times 10^{-6}$ eV$^{2}, \sin^{2} 2\theta = 0.0075.$ 

Just-so vacuum solution(JS): 
$\delta^{2} = 3 \times 10^{-10}$ eV$^{2}, \sin^{2} 2\theta \simeq 1$. 

\subsection{The Complication In Observed Events}

In what follows, an initial flux described
 by a Fermi-Dirac spectrum with zero chemical potential will be assumed.  The
detailed time evolution during the cooling phase has been ignored,
while the averaged magnitudes and effective temperatures of
$\nu(\overline{\nu})$ flux are used instead
~\cite{kh:92}.  The event numbers at the detectors for the neutrino of 
type $\ell$
are estimated by
    \begin{equation}
       N_{\ell}=\frac{Z\,L_{\ell}}{4\pi\, D^{2}} \int dE_{\nu}\, 
n_{\ell}(E_{\nu},T_{\ell})
     \, \sigma_{\ell}(E_{\nu})\,P_{\ell}(E_{\nu}).
   \end{equation}
Here $Z$ is the number of targets in the detector, $L_{\ell}$ is the initial
number of $\nu_{\ell}$, $D$ is the distance between the supernova and the earth,
$\sigma_{\ell}$(E$_{\nu}$) is the cross section for the corresponding reaction,
$P_{\ell}(E_{\nu})$ is the surviving probability for $\nu_{\ell}$,
$T_{\ell}$ is the temperature for $\nu_{\ell}$, and
    \begin{equation}
    n_{\ell}(E_{\nu},T_{\ell}) \simeq 0.5546 \frac{E_{\nu}^2}{T_{\ell}^3 \,
      [1+exp(\frac{E_{\nu}}{T_{\ell}})]}.
     \end{equation}
In evaluating the surviving probability, the electron number
per nucleon is assumed to remain a constant ($Y_{e} \approx 0.42$) and
the density profile outside the neutrino-sphere ($r \geq 10^{7}$ cm)
is described by
the power-law $\rho \sim r^{-3}$.

To grasp the picture on how uncertainties
in neutrino temperatures could affect the
interpretation of observed events, 
we may tentatively assume $T_{{\nu}_{e}}$=3 MeV, $T_{{\nu}_{x}}=
T_{\overline{\nu}_{x}}$= 6 MeV ($x= \mu, \tau$),
 and compare outcomes from $T_{\overline{\nu}_{e}}$=3 MeV
and  $T_{\overline{\nu}_{e}}$=4.5 MeV.  At
Super-Kamiokande~\cite{fukuda:1998}, contributions from  
 the inverse beta decay $\overline{\nu}_{e}+p \rightarrow e^{+}+n $ 
 predominate due to the high cross section. 
Events from other interactions: $\nu_{\ell} (\overline{\nu}_{\ell}) + e^{-},
(\ell=e,\mu,\tau)$~\cite{arafune:87} 
and $\nu_{e}( \overline{\nu}_{e})+  ^{16}O$ 
~\cite{haxton:87}, will also be
included in our calculations although these events accumulate up to less
 than 5\% of the $\overline{\nu}_{e}+p$ events.  The threshold
energy is taken to be 5 MeV and the detector efficiency
is assumed to be 100\%.  For 32 kton of water, one expects roughly
$\sim 10^{4}$ neutrino events for a type II supernova at the
center of our galaxy ($\sim$ 10 kpc away).

Since the cross section for
$\overline{\nu}_{e} + p$ is proportional to $E_{\nu}^{2}$,
and $E_{\nu}\simeq 3.1 T_{\nu}$ for the Fermi-Dirac distribution, 
a larger temperature gap between $\overline{\nu}_{e}$ and
$\overline{\nu}_{x}$ would cause a
more severely distorted spectra from the original one. 
Hence the difference between
$T_{\overline{\nu}_{e}}$ and $T_{\overline{\nu}_{x}}$ determines
to what extend the events are enhanced by oscillation.

For the direct masses where $m_{\nu_{\tau}} \gg m_{\nu_{\mu}} > m_{\nu_{e}}$, 
possible results of the ratio $OSC/NO$ ($OSC$ indicates oscillation, 
and $NO$ indicates the case of
no oscillation) using specific input parameters LA, SA, 
and JS are shown in Figure 2. 
The curves representing LA and SA are due to MSW effects of the  
$\nu$-type events and the vacuum oscillation of the $\overline{\nu}$-type
events, while the JS curve is due to vacuum oscillations of both $\nu$- and
$\overline{\nu}$-type events.
One observes that  
JS parameters could raise event
numbers most effectively, an increase of 
$\sim$ 55\% is possible at
$T_{\overline{\nu}_{e}}\simeq $3 MeV ( $\frac{T_{\overline{\nu}_{e}}}
{T_{\overline{\nu}_{x}}} \simeq 0.5$ ).
The enhancement decreases as $T_{\overline{\nu}_{e}}$
approaches $T_{\overline{\nu}_{x}}$. 
Near a particular point where 
$T_{\overline{\nu}_{e}} \simeq T_{\overline{\nu}_{x}}$, 
the conversion of $\overline{\nu}_{e}$ to $\overline{\nu}_{x}$ would
not alter the original $\overline{\nu}_{e}$ spectrum, 
all the
scenarios yield $OSC/NO \simeq 1$ and are indistinguishable among each other. 

The complication arises from the fact that if, for instance,
$OSC/NO=1.3$ is observed, this observation is
then either due to LA parameters at 
$T_{\overline{\nu}_{e}} \simeq 3.6$ eV or the JS parameters  
at $T_{\overline{\nu}_{e}} \simeq 4$ eV.   
The uncertainties in neutrino temperatures would therefore render a
wide range of predictions at the detectors. 
Informations such as
the clues for oscillation and neutrino
parameters
would be hard to understand or even lost due to this
complication.

\section{General Consequences From The Direct Masses}
\subsection{Super-Kamiokande}

For the observation of SN neutrinos, 
the extremely distinct time structures between the neutronization burst 
and the beginning of thermal emission  
would allow a clear separation at the $H_{2}O$ Cherenkov detector. 
These two groups of events are discussed separately.

The spectral shape and the total energy of
$\nu_{e}$ from the early pre-bounce burst is still poorly known.
For the purpose of qualitative discussion, the spectrum is arbitrarily
chosen to be the same as
that of thermal $\nu_{e}$ (Fermi-Dirac) with the same mean energy and
a total of 5\% the binding energy of a typical neutron star ($E_{b}$).
During this early phase, one expects to observe the forward
directional events due to elastic scattering  $\nu_{e} +e^{-}$
and the backward events from $\nu_{e} +  ^{16}O$.
These neutronization events are
summarized in Table I.  We note that
the forward events are
relatively insensitive to the uncertainty in
$\nu_{e}$ temperature.  The oscillation signature manifests itself through the
drastically reduced forward events as compared to the original one, 
although practically the separation
among LA, SA, and JS using events observed during this early
phase is difficult.

The backward events on the other hand, 
are more sensitive to $T_{\nu_{e}}$.  The difficulty associated with
the backward events comes from the extremely small numbers.
If $T_{\nu_{e}}= 3$ MeV and the total neutrino energy at this stage is down to
$\sim 1\%$ of $E_{b}$, the backward events are practically unobservable.
Because of the rapidly increased cross section for  
$\nu_{e} +  ^{16}O$ at higher energy:
$\sigma \sim (E-E_{th})^{2}$ , 
the situation could be improved if $T_{\nu_{e}}$
is higher or if the neutrinos emitted
during this phase have larger energy partition, which is quite
model-dependent.  
Unlike the backward events, the forward event numbers are
roughly in the order of 10 even if
$T_{\nu_{e}}=$ 3 MeV and the $\nu_{e}$ flux takes away as low as
$\sim 1\%$ of $E_{b}$.

By using the numerous 
$e^{+}$ emitted from the inverse beta decay, 
the distorted $\overline{\nu}_{e}$ spectrum
would be
determined with better statistics. 
To account for the uncertainties in $T_{\nu_{e}}$ and
$T_{\nu_{x}}$, we may let
$T_{\nu_{x}}=\alpha$
$T_{\overline{\nu}_{e}}$, $T_{\nu_{e}}=\beta$  $T_{\overline{\nu}_{e}}$
and compare the outcomes for $T_{\overline{\nu}_{e}}$= 4 MeV, 5 MeV, and 6 MeV.
 The parameters $\alpha$ and $\beta$ are allowed to vary within
$1.4\leq \alpha \leq 1.8$ and $0.6 \leq \beta \leq 1$ to
roughly include the temperature ranges given by current models.
 Expected ranges for the ratios
$R \equiv I/F$  are summarized in Table II, 
 where $I$ includes events from the inverse beta decay
and the neutrino interactions with oxygen, 
$F$ represents the forward scattering events. 
For $T_{\overline{\nu}_{e}}= 4$ MeV, there are two overlapped
areas in $R$: between NO, SA and between JS, LA.
The same overlapping structure
 remains for $T_{\overline{\nu}_{e}}= 5$ MeV and
$T_{\overline{\nu}_{e}}= 6$ MeV.
Despite the wealthy information conveyed through the
$e^{+}$ spectrum at Super-Kamiokande, from Table II 
it seems unlikely that a clear separation among input
parameters could be achieved using the otherwise model-independent quantity
$R$ unless the uncertainties in neutrino temperatures are reduced
significantly.

\subsection{SNO}

The neutral current$(NC)$ breakup reactions of deuterium in SNO~\cite{sno:97}
are flavor-blind for neutrinos:
  \begin{equation}
   \nu_{\ell} + d \rightarrow n+p+\nu_{\ell}, \, (E_{th}=2.22 MeV)\\
  \end{equation}
  \begin{equation}
\overline{\nu}_{\ell} + d \rightarrow n+p+\overline{\nu}_{\ell}, \, 
(E_{th}=2.22 MeV)\\
   \end{equation}
where $\ell = e, \nu, \tau$.
  The charged current reactions include two parts,
   \begin{equation}
   CC_{1}:  \nu_{e} + d \rightarrow p+p+e^{-}, \, (E_{th}=1.44 MeV) \\
   \end{equation}
 \begin{equation}
    CC_{2}: \overline{\nu}_{e} + d \rightarrow n+n+e^{+}, \,
     (E_{th}=4.03 MeV). \\
   \end{equation}

With 1 kton of $D_{2}O$ and a threshold energy $\sim$ 5 MeV
( 100\% detection efficiency assumed), both $CC_{1}$ and $CC_{2}$ should
roughly yield event numbers in the order of $10^{2}$.  The ratios
$r_{1} = \frac{NC}{CC_{1}}$ at SNO seem
 to provide a solution as to how a particular set of parameter could
be singled out, as will be shown below. 

One may arbitrarily fix $T_{\nu_{e}}$
and parameterize other temperatures in a
similar way: let $T_{\nu_{x}}=\lambda T_{\nu_{e}}$ and
allow an uncertainty in $T_{\overline{\nu}_{e}}$ as well:
$T_{\overline{\nu}_{e}}=\eta T_{\nu_{e}}$, with
$1.8 \leq \lambda \leq 2.6$ and $1.1 \leq \eta \leq 1.7$.
 The ratio $r_{1}=\frac{NC}{CC_{1}}$
for $T_{\nu_{e}}=3,4$ and 5 MeV are listed in Table III. 
We found that even if the uncertainties
in $T_{\nu_{x}}$ and $T_{\overline{\nu}_{e}}$ may complicate
the interpretation of observed events, each of the
candidates gives rise to a distinct region in $\frac{NC}{CC_{1}}$.
In practical, if
 uncertainties in $\lambda$ and $\eta$ can be reduced
in the future,
it would enable a smaller spread in each $r_{1}$ 
for a better separation.

\section{Consequences From The Inverted Masses}
\subsection{Vacuum oscillation versus MSW effect}

In the light of MSW effect, distinctions between direct and
inverted masses would most likely appear in the observed
neutrino spectra.
If the just-so vacuum oscillation is favored over the
MSW oscillations as solution to the SNP, both
direct ($m_{\nu_{\tau}} \gg m_{\nu_{\mu}} > m_{\nu_{e}}$) and inverted-mass
schemes ($m_{\nu_{e}} > m_{\nu_{\mu}} \gg m_{\nu_{\tau}}$)
are allowed since $\nu_{e}$ flux can
also be converted to $\nu_{x}$ through the vacuum oscillation if neutrino
masses are inverted.  The $\overline{\nu}_{e}$ flux on the contrary, would
go through the MSW resonance if the mass hierarchy is inverted.  This
conversion would presumably enlarge the $\overline{\nu}_{e}$-type
event rates effectively at Super-Kamiokande.
Without conflicting current
solar and atmospheric neutrino data, we would focus on the JS
parameters: 
$-\delta^{2} \sim 10^{-10}$eV$^{2}$ and large $\theta$, for a  
further investigation under the inverted mass scheme
$m_{\nu_{e}} > m_{\nu_{\mu}} \gg m_{\nu_{\tau}}$.

The possible SN $\overline{\nu}_{e}$ spectra are shown in
Figure 3.  Curve $A$ is the original
$\overline{\nu}_{e}$ spectrum and curve $B$ 
represents the distorted one through the just-so vacuum oscillation under
the direct-mass scheme while
curve $C$ is obtained from the MSW conversion under the inverted-mass scheme.
Curves $B$ and $C$ 
nearly overlap,
 implying that the matter effect
is not as prominent as expected, and that the MSW effect under the 
inverted-mass hierarchy   
is almost identical to the vacuum oscillation 
under the direct-mass hierarchy for $\overline{\nu}_{e}$ at
this particular region of parameter space.  Furthermore,
the extremely small observable difference at the detectors
would make the
identification between the two mass patterns very difficult.
 The reason becomes clear if the
required conditions for a MSW resonance and
an adiabatic transition to occur are both considered~\cite{totani:96}: 
A density profile $\rho \sim r^{-3}$
would yield $|\delta^{2}| \sim 10^{-8}-10^{5}$ eV$^{2}$
relevant to the MSW oscillation in the
supernova. 
This mass scale is much larger
than the mass scale of JS parameters
($|\delta^{2}| \sim 10^{-10}$ eV$^{2}$).  Therefore a very effective
conversion of $\overline{\nu}_{e}$ to $\overline{\nu}_{x}$ in a
supernova is unlikely for either direct or inverted masses if
the JS parameters are applied.
The strong conversion of $\overline{\nu}_{e}$ to $\overline{\nu}_{x}$
is actually disfavored by some analyses based on the SN 1987A 
data~\cite{smirnov:94}.
If either LA or SA MSW conversion
is favored over the just-so vacuum oscillation,
the case for the inverted hierarchy
$m_{\nu_{e}} > m_{\nu_{\mu}} \gg m_{\nu_{\tau}}$
would then become shaky or can even be ruled out.

\subsection{Super-Kamiokande and SNO}

An alternative approach might shed some clues to the inverted-mass
scheme and its outcomes.  We may characterize
consequences for the oscillation $\overline{\nu}_{e} \leftrightarrow
\overline{\nu}_{x}$
by the surviving probability of
$\overline{\nu}_{e}$ in
three limit cases: $P(\overline{\nu}_{e} \rightarrow \overline{\nu}_{e})
\sim 1, P(\overline{\nu}_{e} \rightarrow \overline{\nu}_{e}) \sim \frac{1}{2}$, 
and $P(\overline{\nu}_{e} \rightarrow \overline{\nu}_{e})\ll 1$.
The case $P(\overline{\nu}_{e} \rightarrow \overline{\nu}_{e}) \sim $ 1 
indicates that no conversion occurs among
$\overline{\nu}_{e}$ and $\overline{\nu}_{x}$,
which is equivalent to the outcome of having massless neutrinos,
and the mass pattern would then unlikely be the main issue.
The JS parameters, as already discussed, yield
$P(\overline{\nu}_{e} \rightarrow \overline{\nu}_{e}) \sim \frac{1}{2}$
for the MSW conversion (or equivalently, the vacuum oscillation
under the direct-mass scheme).  One is therefore motivated to further
study the consequences for a complete conversion of
$\overline{\nu}_{e}$ flux to $\overline{\nu}_{x}$, where 
$P(\overline{\nu}_{e} \rightarrow \overline{\nu}_{e}) \ll 1$.
As pointed out by Totani \emph{et al.}~\cite{totani:96},
due to statistical
uncertainties in experiments and the inconsistency among current analyses,
one cannot completely exclude the possibility of full conversion.
We may tentatively
neglect details of the physical conditions and
parameters that required for a complete conversion
to occur, and assume that the probability 
$P(\overline{\nu}_{e} \rightarrow \overline{\nu}_{e})$ remains 
approximately a
constant within interested range of the neutrino energy.

To reasonably account for the contributions from $\nu_{e}$ and $\nu_{x}$
fluxes to the total events when a complete conversion occurs in 
the anti-neutrinos sector 
, one may first consider 
the contours of $OSC/NO$ 
for events from the inverse beta decay only.  Under the inverted-mass scheme, 
a wide range of
$-\delta^{2}$ and $\theta$ are shown in
Figure 4.
Given $T_{\overline{\nu}_{e}}=$ 4.5 MeV,
          $T_{\nu_{e}}$=3 MeV and
           $T_{\nu_{x}}=
                T_{\overline{\nu}_{x}}$=6 MeV,  
the JS parameters ($|\delta^{2}| \sim 10^{-10}$ eV$^{2}$
and large $\theta$) roughly result in a 20\% increment  
to the event number. 
After a full conversion of the anti-neutrinos in which 
$P(\overline{\nu}_{e} \rightarrow \overline{\nu}_{e}) \ll 1$, 
one would expect to 
observe a
sizable increase in the ratio $OSC/NO$.  Therefore a
larger $|\delta^{2}|$ and a smaller $\tan^{2}2\theta$, at least
several orders of magnitude, 
are required for a full conversion to occur. 
The smallness of $\theta$, alone with the small $\phi$,
fix the surviving probability of $\nu_{e}$ 
very close to unity.  Hence, a full conversion of $\overline{\nu}_{e}$ to
$\overline{\nu}_{x}$ would be accompanied by nearly unchanged
$\nu_{e}$ and $\nu_{x}$ fluxes:   
$P(\nu_{e} \rightarrow \nu_{e} ) \sim 1$. 
The expected ratios $R\equiv I/F$ at the Super-Kamiokande are
listed in Table IV.
Despite the better statistic provided by the $\overline{\nu}_{e}$-type events 
at Super-Kamiokande, 
the detection is however not unique to the
$\overline{\nu}_{e}$ flux, the uncertainties in 
$T_{{\nu}_{e}}$ and $T_{{\nu}_{x}}$   
therefore make the separation between  
$P(\overline{\nu}_{e} \rightarrow \overline{\nu}_{e}) \ll 1$ 
(complete conversion) and
$P(\overline{\nu}_{e} \rightarrow \overline{\nu}_{e}) 
\sim \frac{1}{2}$ difficult. 

At the SNO detector, the charged-current channel
$\overline{\nu}_{e} + d \rightarrow n + n + e^{+}$
is unique to $\overline{\nu}_{e}$.  This channel can be distinguished from 
the neutral-current events
and the other charged-current channel induced by
$\nu_{e}$.  
Therefore the measurement of $\overline{\nu}_{e} + d $ events
at SNO should be sensitive to the full conversion of $\overline{\nu}_{e}$
to $\overline{\nu}_{x}$.
Ratios of the neutral-current events to the charged-current events
$\overline{\nu}_{e} + d $, denoted as $\frac{NC}{CC_{2}}$,
are shown in Table V.  We also present values of the same ratio
under the direct-mass scheme in Table VI as a comparison.

We observe that for a particular $T_{\nu_{e}}$,
both direct and inverted
schemes yield a nearly identical range of $\frac{NC}{CC_{2}}$ if the JS
parameters are applied, this verifies a previous argument.  The valuable
message from this ratio is that for a particular $T_{\nu_{e}}$,
a complete conversion of the
$\overline{\nu}_{e}$ flux through MSW resonance represents a unique range of
$\frac{NC}{CC_{2}}$ as compared to other scenarios, including that of the
direct masses.

Typical $\frac{NC}{CC_{2}}$ contours for
inverted masses
are shown in Figure 5, where $T_{\nu_{e}}$= 3 MeV,
$T_{\overline{\nu}_{e}}$= 4.5 MeV,
and $T_{\nu_{x}}= T_{\overline{\nu}_{x}}$= 6 MeV are assumed.
Since the neutral-current
reaction is blind to the oscillation, 
a complete swap of
$\overline{\nu}_{e}$ and $\overline{\nu}_{x}$
fluxes would definitely yield smaller $\frac{NC}{CC_{2}}$. 
Calculations show that $\frac{NC}{CC_{2}} \sim 2.95$ 
for a complete conversion and  
indicate that $-\delta^{2}>10^{-2}$ eV$^{2}$
and $\tan^{2}2\theta<10^{-2}$ are required for
a near complete conversion to occur.

Tables V and VI suggest that for the detection of SN neutrinos,
the direct and inverted masses could be
 distinguishable if a nearly complete conversion of $\overline{\nu}_{e}$
to $\overline{\nu}_{x}$ occurs, which yields a low $\frac{NC}{CC_{2}}$
and signals the existence of inverted-mass pattern.
Since the MSW effects become important for supernova neutrinos at
$10^{-8}<\Delta m^{2}<10^{5}$ eV$^{2}$, a
future supernova would provide a test ground for 
 $-\delta^{2} > 10^{-2}$ eV$^{2}$.
If the nearly full
conversion $\overline{\nu}_{e} \leftrightarrow \overline{\nu}_{x}$ is
observed in the SN neutrino flux,
the consequences
may have certain implications in that the required parameter spaces
for a full conversion are obviously disfavored by current
solar neutrino data, while 
the future solar and atmospheric observations may not severely change the mass
scales required to explain the solar and the atmospheric neutrino deficits.  

\section{Discussions and Conclusions}

Under the constraints of mass scale from solar
and atmospheric neutrinos, the $3-\nu$ scenario naturally
leads to four possible hierarchies
(one direct and three inverted):
$(1) m_{\nu_{\tau}} \gg m_{\nu_{\mu}} > m_{\nu_{e}},
(2) m_{\nu_{\mu}} > m_{\nu_{e}} \gg m_{\nu_{\tau}}
(3) m_{\nu_{e}} > m_{\nu_{\mu}} \gg m_{\nu_{\tau}},
(4) m_{\nu_{\tau}} \gg m_{\nu_{e}} > m_{\nu_{\mu}}.$
Case 1 is the normal, direct mass scheme.  For 
Case 2, $|\delta^{2}|$ in the mass scale of the MSW
solution has been discussed~\cite{dighe:99}.  In our analysis we have
applied Case 3, in which the mass scale of
$|\delta^{2}|$ is suitable for the vacuum
solution of SNP ($10^{-10}$ eV$^{2}$).
Case 3 and Case 2
become equivalent if $|\delta^{2}| \sim 10^{-10}$ eV$^{2}$
since the MSW and the vacuum
oscillations for the $\overline{\nu}_{e}$ flux would be nearly 
identical at this mass scale, as shown
in Section IV. 
For Case 4 to survive, $|\delta^{2}|$
needs to be in the order
of $10^{-10}$ eV$^{2}$ for the vacuum solution to apply.  Therefore,
consequences for Case 4
and Case 1 become equivalent in the detection of
supernova neutrinos if
 $|\delta^{2}| \sim 10^{-10}$ eV$^{2}$.

In this work, responses at both Super-Kamiokande and SNO detectors
to neutrino fluxes coming from a supernova are
studied under the  consideration of uncertainties in
neutrino temperatures.
In particular, some phenomenological consequences for
direct-mass and
inverted-mass
 patterns  of neutrinos are compared.
We may summarize our results as follows.
(a) Uncertainties in neutrino temperatures can allow
various interpretations of neutrino parameters.  We have shown
this through the expected outcomes at Super-Kamiokande for SN neutrinos. 
(b) The three candidates LA, SA, and JS manifest 
differently in the ratio $\frac{NC}{CC_{1}}$ at SNO even if
the uncertainties in neutrino temperature are allowed.  Future detection
of SN neutrinos at SNO would be able to single out favored  
mass and mixing parameters from the three candidates.
(c) In addition to the direct-mass pattern,
the inverted-mass scenario
$m_{\nu_{e}} > m_{\nu_{\mu}} \gg m_{\nu_{\tau}}$ is investigated  
since it can allow the vacuum solution to the solar neutrino problem. 
By using the event ratio $\frac{NC}{CC_{2}}$
in SNO, the  
direct-mass($m_{\nu_{\tau}} \gg m_{\nu_{\mu}} > m_{\nu_{e}}$) and 
the inverted-mass($m_{\nu_{e}} > m_{\nu_{\mu}} \gg m_{\nu_{\tau}}$)
could be distinguished if a nearly complete 
$\overline{\nu}_{e} \leftrightarrow \overline{\nu}_{x}$
conversion occurs in the
anti-neutrino section. 
\acknowledgments S.C. would like to thank Nien-Po Chen and Sadek Mansour
for suggestions in preparing the manuscript.  
T. K. is supported in part by the DOE, Grant no.
     DE-FG02-91ER40681.



  \begin{table}
 \begin{center}
 \begin{tabular}{|c||cc|cc||cc|}
 \multicolumn{1}{c}{} &
  \multicolumn{2}{c}{$T_{\nu_{e}}$ =3 MeV} &
   \multicolumn{2}{c}{$T_{\nu_{e}}$ =4 MeV} &
     \multicolumn{2}{c}{$T_{\nu_{e}}$ =5 MeV} \\  \hline
          &  backward   &  forward  & backward  &  forward  &  backward  &  forward  \\  \hline
      NO  &  7   & 68  &20  & 71  & 38  &  72 \\  \hline
      LA  &  2   & 26  & 5  & 27  & 10   & 29  \\  \hline
      SA  &  5   & 41  & 14 & 46  & 28  & 49  \\  \hline
      JS  &  3   & 38  & 9  & 39  &19   & 42  \\  \hline
    \end{tabular}
   \caption{Total expected backward and 
      forward events at
   Super-Kamiokande for the neutronization
        burst of neutrinos from a typical supernova $\sim 10$ kpc away.}
   \end{center}
 \end{table}
 \begin{table}
 \begin{center}
 \begin{tabular}{|c||c||c||c|}  \hline
     & $T_{\overline{\nu}_{e}}$= 4 MeV &
       $T_{\overline{\nu}_{e}}$= 5 MeV &
       $T_{\overline{\nu}_{e}}$= 6 MeV  \\ \hline
      NO   & 15.9 - 16.9  &  20.0 - 20.9  &  23.9 - 25.0  \\ \hline
      LA   & 18.3 - 20.9  &  22.7 - 25.4  &  26.7 - 29.4  \\ \hline
      SA   & 16.4 - 18.1  &  20.9 - 23.4  &  24.8 - 27.7  \\ \hline
      JS   & 19.1 - 22.5  &  23.7 - 26.8  &  27.2 - 29.6  \\ \hline
     \end{tabular}
    \caption{
   The ratio $R\equiv I/F$ in Super-Kamiokande for thermally
    emitted neutrinos.  Here the mass hierarchy is direct.  
The uncertainties in $T_{\nu_{e}}$ and
      $T_{\nu_{x}} (T_{\overline{\nu}_{x}})$ give rise to the  
        spread in $R$.}
  \end{center}
 \end{table}
 \begin{table}
 \begin{center}
 \begin{tabular}{|c||c||c||c|}  \hline
     & $T_{\nu_{e}}$= 3 MeV &
       $T_{\nu_{e}}$= 4 MeV &
       $T_{\nu_{e}}$= 5 MeV  \\ \hline
      NO   & 5.4 - 9.0   &  5.3 - 7.9  &  5.1 - 6.5   \\ \hline
      LA   & 2.6 - 2.8   &  2.7 - 2.9  &  2.7 - 2.9      \\ \hline
      SA   & 3.9 - 5.1   &  4.0 - 5.1   &  4.0 - 4.9  \\ \hline
      JS   & 3.1 - 3.5   &  3.2 - 3.5   &  3.2 - 3.5  \\ \hline
     \end{tabular}
    \caption{ $r_{1}=\frac{NC}{CC_{1}}$,
          the ratios of neutral-current event numbers to charged-current
       event numbers ($\nu_{e}+d \rightarrow p+p+e^{-}$) in SNO for
        thermally emitted neutrinos.  
        Here the mass hierarchy is direct. }
   \end{center}
 \end{table}
\begin{table}
 \begin{center}
 \begin{tabular}{|c||c||c||c|}  \hline
     & $T_{\overline{\nu}_{e}}$= 4 MeV &
       $T_{\overline{\nu}_{e}}$= 5 MeV &
       $T_{\overline{\nu}_{e}}$= 6 MeV  \\ \hline
      JS(INVERTED)   & 19.3 - 22.8  &  23.6 - 26.5  &  27.3 - 29.6  \\ \hline
      COMPLETE   & 22.1 - 28.3  &  26.3 - 30.6 &  29.2 - 30.9  \\ \hline
     \end{tabular}
    \caption{The ratio $R\equiv I/F$ at Super-Kamiokande
      under the inverted-mass scheme.}
  \end{center}
 \end{table}
\begin{table}
 \begin{center}
 \begin{tabular}{|c||c||c||c|}  \hline
     & $T_{\nu_{e}}$= 3 MeV &
       $T_{\nu_{e}}$= 4 MeV &
       $T_{\nu_{e}}$= 5 MeV  \\ \hline
      JS & 3.4 - 4.0   &  3.2 - 3.8 &  3.1 - 3.7  \\ \hline
      COMPLETE   & 2.6 - 3.2   &  2.6 - 3.1  & 2.6 - 3.0      \\ \hline
     \end{tabular}
    \caption{$\frac{NC}{CC_{2}}$, the ratios of neutral-current event
             numbers to charged-current event numbers
      ($\overline{\nu}_{e}+d \rightarrow n+n+e^{+}$) in SNO for thermally
          emitted neutrinos.  Here neutrino masses are inverted.  This
               table can be compared with Table VI where neutrino 
                    masses are direct. 
        A complete conversion yields the lowest value of $\frac{NC}{CC_{2}}$
             for a given
             $T_{\nu_{e}}$.
                  }
   \end{center}
 \end{table}
\begin{table}
 \begin{center}
 \begin{tabular}{|c||c||c||c|}  \hline
     & $T_{\nu_{e}}$= 3 MeV &
       $T_{\nu_{e}}$= 4 MeV &
       $T_{\nu_{e}}$= 5 MeV  \\ \hline
      NO   & 3.5 - 11.4   &  3.4 - 8.8 &  3.2 - 6.8  \\ \hline
      LA   & 3.4 - 4.7   &  3.2 - 4.4  &  3.2 - 4.1      \\ \hline
      SA   & 3.5 - 9.4   & 3.3 - 7.7 &  3.2 - 6.2  \\ \hline
      JS   & 3.4 - 4.1   &  3.2 - 3.9   &  3.1 - 3.7  \\ \hline
     \end{tabular}
    \caption{$\frac{NC}{CC_{2}}$ for the thermal emission of neutrinos in SNO,
        with direct masses. }
   \end{center}
 \end{table}

 \begin{figure} 
\caption{$P(\nu_{e} \rightarrow \nu_{e})$ for SN neutrinos at 
(a) $\delta^{2}/E=10^{-5}$ eV$^{2}$/MeV,
        (b) $\delta^{2}/E=10^{-7}$ eV$^{2}$/MeV, 
     (c) $\delta^{2}/E=10^{-9}$ eV$^{2}$/MeV, and
         (d) $\delta^{2}/E=10^{-11}$ eV$^{2}$/MeV.  
Here $\delta^{2}$ is the mass-squared difference in eV$^{2}$ and $E$ is neutrino energy in MeV.} 
\label{fig1.ps}
\end{figure}
 \begin{figure} 
\caption{Predicted curves of the ratio
    $OSC/NO$ in Super-Kamiokande
      for the direct masses.  We have defined
     $Y \equiv OSC/NO , X \equiv \frac{T_{\overline{\nu}_{e}}}
     {T_{\overline{\nu}_{x}}}$.  Here
         $T_{\nu_{e}}=3$ MeV, $T_{\overline{\nu}_{x}}$=6 MeV are assumed,
      and $T_{\overline{\nu}_{e}}$ are allowed to vary from 3 MeV to 6 MeV
    as indicated by $X \equiv \frac{T_{\overline{\nu}_{e}}}
     {T_{\overline{\nu}_{x}}}$.  
 } 
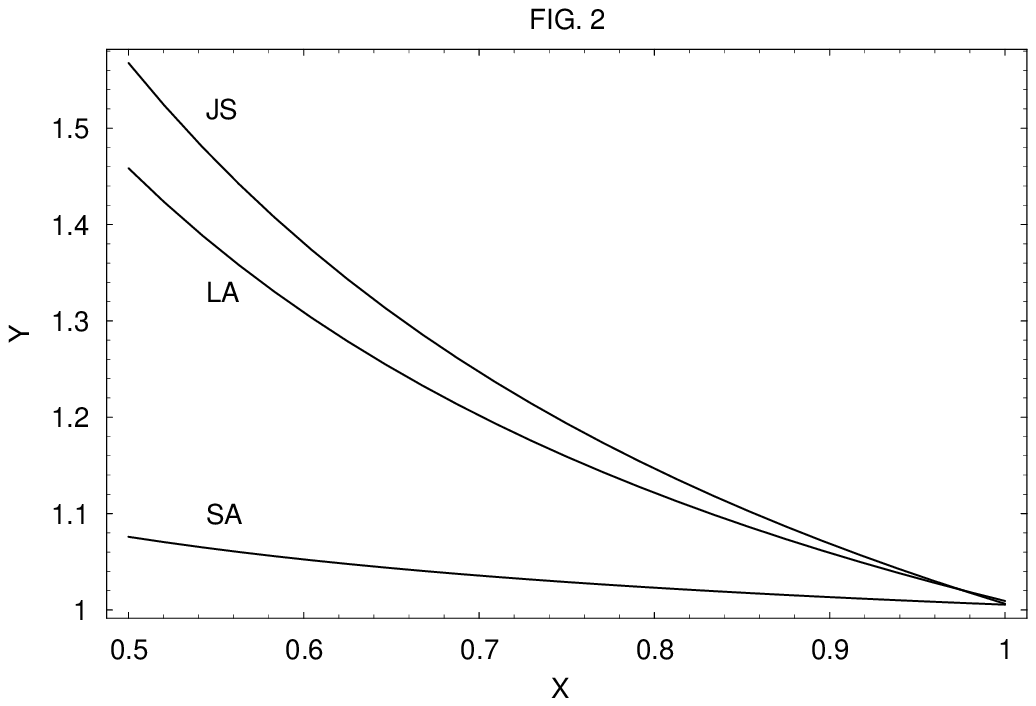
\label{fig2.ps}
    \end{figure}
\begin{figure}
\caption{Expected $\overline{\nu}_{e}$ spectra from a supernova.
    Curve $A$ represents the
    original $\overline{\nu}_{e}$ without any conversion;
        curve $B$ is the one distorted
    by the vacuum oscillation under direct-mass hierarchy
       while $C$ is the expected curve
   for $\overline{\nu}_{e}$ after MSW conversion, with the same parameters applied
    in $B$ but under inverted-mass scheme.  $B$ and $C$ are nearly indistinguishable.}
  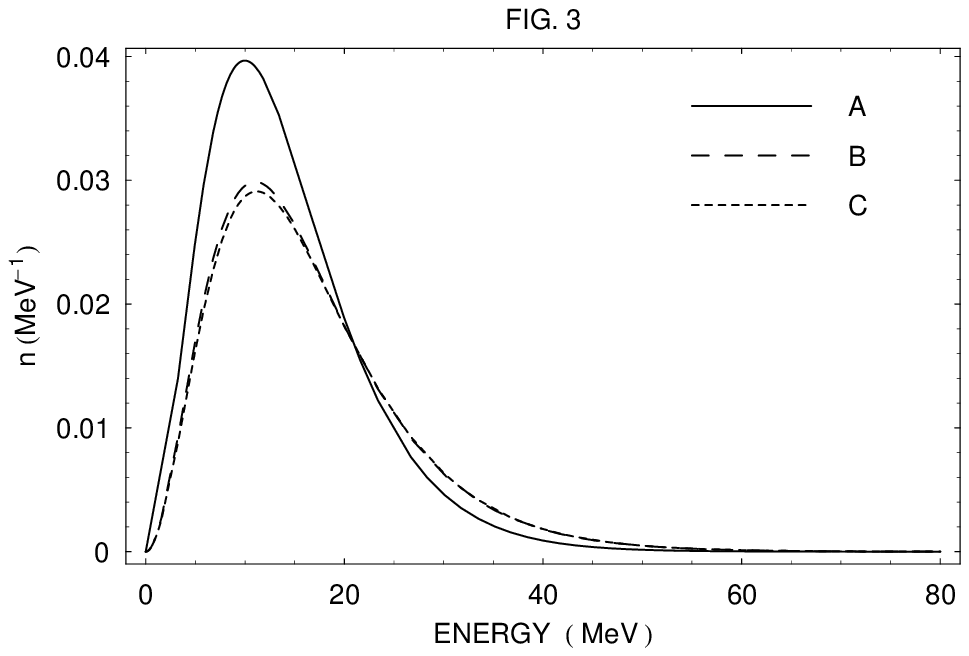
\label{fig3.ps}
    \end{figure} 

\begin{figure}
\caption{The contour plot of $OSC/NO$ from the events of inverse beta decay. 
     The inverted masses are assumed.  Here we have set 
          $T_{\overline{\nu}_{e}}=$ 4.5 MeV,  
          $T_{\nu_{e}}$=3 MeV and
           $T_{\nu_{x}}=
                T_{\overline{\nu}_{x}}$=6 MeV.}
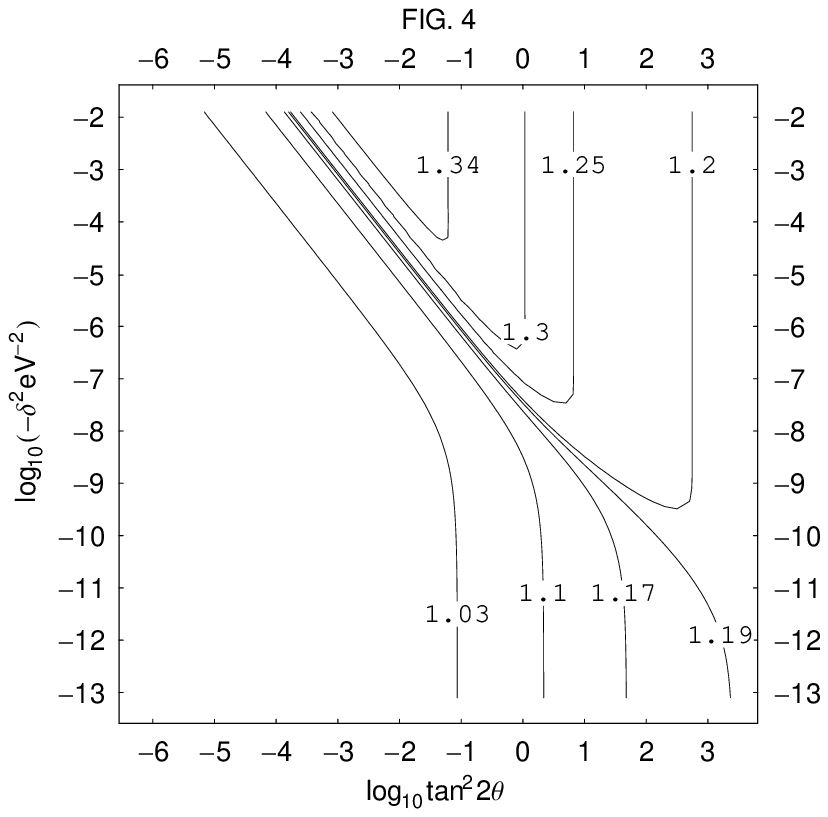
\label{fig4.ps}
 \end{figure}

\begin{figure} 
  \caption{Typical $NC/CC_{2}$ contours in SNO under the inverted-mass scheme.
     Here $T_{\nu_{e}}$=3 MeV, $T_{\overline{\nu}_{e}}$=4.5 MeV and
    $T_{\nu_{x}}=T_{\overline{\nu}_{x}}$=6 MeV are assumed.
    A complete conversion of the $\overline{\nu}_{e}$ flux
   to $\overline{\nu}_{x}$ flux would yield $NC/CC_{2} \sim 2.95$,
which occurs at $|-\delta^{2}| > 10^{-2}$ eV$^{2}$.} 
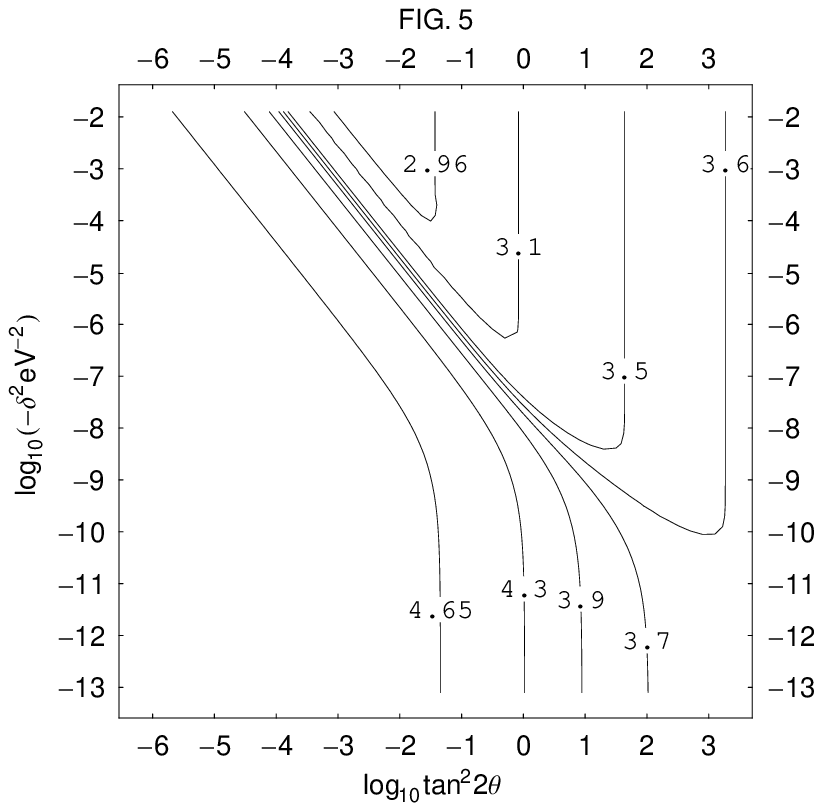
\label{fig5.ps}
    \end{figure}

\end{document}